\documentclass[10pt,journal]{IEEEtran}

\usepackage{cite}
\usepackage{amsmath,amssymb,amsfonts}
\usepackage{array}
\usepackage{booktabs}
\usepackage{graphicx}
\usepackage{algorithm}
\usepackage{algorithmic}
\usepackage{textcomp}
\usepackage{xcolor}
\usepackage{url}

\begin{document}

\title{CoSimRec: Measuring Coordinated-Content Penetration in Recommender Feedback Loops}

\author{Nan~Li, Jiahong~Shao, and Jiuyang~Lyu%
\thanks{Nan Li, Jiahong Shao, and Jiuyang Lyu are with the State Key Laboratory of Media Convergence and Communication, Communication University of China, Beijing, China (e-mail: linan.tech@gmail.com; shaojiahong2001@outlook.com; lyujiuyang@cuc.edu.cn).}}

\markboth{Preprint,~2026}%
{Li \MakeLowercase{\textit{et al.}}: CoSimRec: Measuring Coordinated-Content Penetration}

\maketitle

\begin{abstract}
Recommender systems shape which content reaches users, making it important to measure whether coordinated activity gains visibility beyond the accounts that initiate it. Existing robustness evaluations largely focus on static target-rank changes and do not capture how coordinated interactions, recommendation, and user response evolve within a feedback loop. We propose CoSimRec, an offline agent-based evaluation framework that models coordinated accounts, dynamic ranking, controlled non-bot responses, and ranking interventions in a shared closed-loop process. CoSimRec introduces the Algorithmic Penetration Rate (APR) metric family: exposure APR is the primary endpoint, while behavior APR is a response-model-conditional sensitivity measure; both can be compared with matched no-attack baselines. We evaluate CoSimRec on MIND, MovieLens, and LastFM with random, popularity-based, feedback-sensitive, MF, BPR-MF, and BPR-LightGCN recommenders. In a risk-blind primary protocol, random controls show no statistically supported positive penetration, whereas popularity-based and feedback-sensitive ranking produce positive APR-Lift in all six master-worker settings, reaching 0.4702 on LastFM. A nine-target MovieLens 1M LightGCN stress test shows positive mean APR-Lift around 25\% injection in all three target-popularity strata, while no-filler profiles remain near zero. Under these controlled conditions, coordinated inputs reach non-bot recommendation slots, providing evidence of a computational pathway from organized activity to audience-level visibility.
\end{abstract}


\section{Introduction}
\IEEEPARstart{R}{ecommender} systems allocate attention in settings where user behavior and algorithmic ranking continually update each other. News feeds, video platforms, social media timelines, movie services, and music platforms decide which items users see, then reuse the resulting clicks, ratings, shares, or listening traces as future ranking signals. Although recommender-system research has developed strong collaborative filtering, ranking, and graph-based models \cite{ricci2022recommender,koren2009matrix,rendle2009bpr,he2020lightgcn}, deployed systems are not static predictors. They are feedback loops in which exposure changes behavior, behavior changes the data, and the revised data changes later exposure.

From a social-scientific perspective, these systems are also institutions of algorithmic gatekeeping: they translate distributed behavioral traces into unequal opportunities for attention and participation \cite{gillespie2014relevance,napoli2014automated}. A coordinated campaign succeeds not simply when accounts act together, but when platform rules recognize their activity as relevance and redistribute the resulting visibility to other users. This framing makes coordinated-content penetration a sociotechnical question about the coupling of organized action, platform affordances, and audience attention, as well as a recommender-robustness question.

Static shilling and poisoning studies ask whether fake profiles or ratings can move a target item \cite{mobasher2007trustworthy,gunes2014shilling,li2016poisoning}, while feedback-loop studies examine exposure inequality \cite{chaney2018algorithmic,mansoury2020feedback,abdollahpouri2019unfairness}. Social-bot research and agent-based simulation address coordinated behavior and sequential interaction \cite{ferrara2016rise,varol2017online,vosoughi2018spread,park2023generative,wang2025user,zhang2024agent4rec,ie2019recsim,yang2024oasis}, but these settings do not jointly measure whether coordinated inputs pass through recommender updates into non-bot exposure and response. Target-rank improvement alone does not establish such penetration.

\begin{figure*}[t]
\centering
\includegraphics[width=0.97\textwidth]{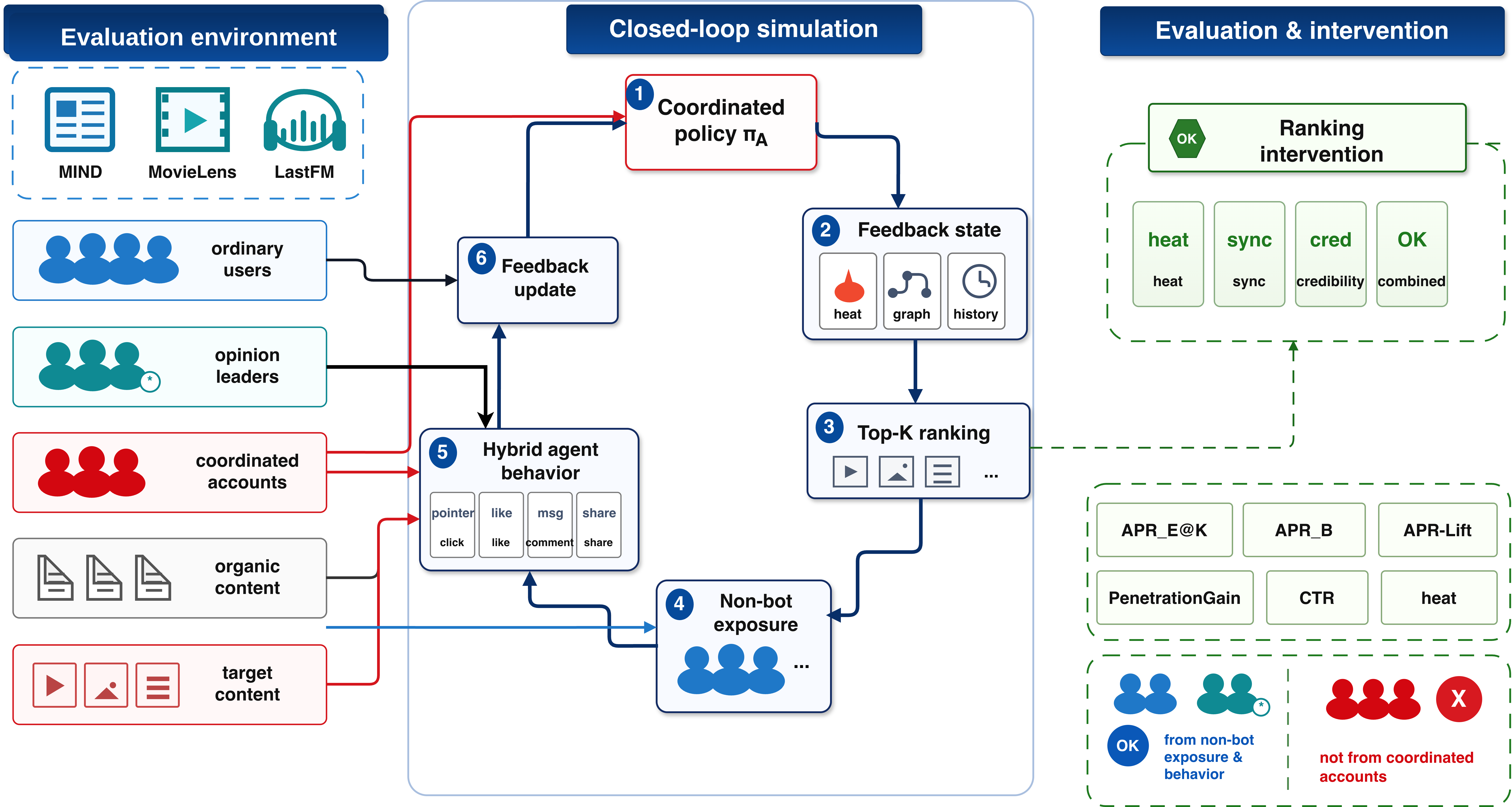}
\caption{CoSimRec measures whether coordinated interactions can pass through recommender feedback and reach non-coordinated users. Coordinated accounts affect the feedback state; the recommender produces Top-K lists; non-bot agents generate behavior; and APR metrics quantify target-content exposure and engagement among ordinary users and opinion leaders.}
\label{fig:framework}
\end{figure*}

CoSimRec measures this conversion in a shared feedback state (Fig.~\ref{fig:framework}). Exposure APR is the fraction of non-bot recommendation slots occupied by target content, and behavior APR is the fraction of non-bot engagement directed to it. Each coordinated run is paired with a no-attack run sharing the seed, users, items, and recommender; APR-Lift thereby removes incidental exposure, while penetration gain normalizes added exposure by the coordinated interaction budget.

We evaluate CoSimRec on MIND \cite{wu2020mind}, MovieLens \cite{harper2015movielens}, and LastFM \cite{grouplens2011hetrec} with random, popularity, feedback, MF, BPR-MF, and BPR-LightGCN recommenders. APR-Lift is near zero under random recommendation but positive for specific feedback-sensitive settings. For BPR-LightGCN, we connect standard profile-injection outcomes to a trained APR loop that retrains after coordinated inputs, serves non-bot Top-K lists, and feeds simulated response into later graph updates.

The paper makes four contributions:
\begin{itemize}
    \item We define coordinated-content penetration as an offline recommender feedback-loop measurement problem, distinct from static attack success and generic social diffusion.
    \item We propose CoSimRec, a reproducible agent-based protocol that combines non-bot response models, coordinated accounts, target content, recommender feedback, and ranking interventions.
    \item We introduce the APR metric family for measuring non-bot target exposure, non-bot target engagement, lift over matched no-attack baselines, and budget-normalized penetration.
    \item Across the evaluated MIND, MovieLens, and LastFM settings, we show that penetration varies by recommender, domain, coordination policy, graph connectivity, and injection scale, and that targeted ranking interventions can reduce APR.
\end{itemize}

\section{Related Work}

\subsection{Recommender Feedback Loops and Robustness}
Ranked exposure becomes training data, which can amplify popularity bias, homogenization, and utility loss \cite{chaney2018algorithmic,mansoury2020feedback,abdollahpouri2019unfairness}. Recommender-robustness research instead studies fake-profile and data-poisoning attacks through manipulated training sets, target ranks, and model-level measures \cite{mobasher2007trustworthy,gunes2014shilling,li2016poisoning}. These perspectives cover different stages of a feedback process: the former characterizes aggregate exposure dynamics, whereas the latter tests target promotion. We report standard ASR and ranking utility alongside closed-loop APR to determine whether promotion transfers into non-bot exposure and response.

\subsection{Agent-Based Simulation and Coordinated Behavior}
Offline simulators support sequential policy comparisons without exposing live users to experimental risk. RecSim provides configurable recommender environments \cite{ie2019recsim}; generative-agent and recommender-agent systems model profiles, histories, memory, and action \cite{park2023generative,wang2025user,zhang2024agent4rec}; and OASIS scales agent interaction to social-media settings \cite{yang2024oasis}. Social-bot and misinformation studies document coordinated effects through platform traces and diffusion analyses \cite{ferrara2016rise,varol2017online,vosoughi2018spread}. CoSimRec combines these strands by holding response specifications and seed schedules fixed while coordination policies and recommender feedback determine target exposure, engagement, and intervention effects.

\subsection{Algorithmic Gatekeeping and Social Influence}
Media and communication research treats ranking algorithms as gatekeepers whose classifications and relevance judgments organize visibility, shape media consumption, and embed institutional choices \cite{gillespie2014relevance,napoli2014automated,nechushtai2019gatekeepers}. This perspective complements robustness analysis by shifting attention from whether a target score changes to who receives the resulting content. It also makes exposure a socially meaningful outcome: recommendation slots are scarce opportunities for attention, although being shown an item does not establish endorsement, persuasion, or harm.

Social-influence experiments show that visible popularity can produce cumulative advantage and inequality \cite{salganik2006experimental}, while exposure-diversity research treats recommender design as a normative choice about what users encounter \cite{helberger2018diversity}. CoSimRec examines a narrower mechanism: it manipulates algorithmic exposure rather than perceived social proof and measures when coordinated signals reach non-bot users under matched conditions.

\section{Problem Formulation}
The evaluation problem is to estimate whether target coordinated content enters non-bot recommendation exposure and engagement during a sequential feedback loop. Let \(U = U_N \cup U_L \cup U_B\) denote the user set, where \(U_N\) contains ordinary users, \(U_L\) contains opinion leaders, and \(U_B\) contains coordinated accounts. Let \(C = C_O \cup C_A\) denote the content set, where \(C_O\) contains organic content and \(C_A\) contains target coordinated content. The simulation proceeds over discrete time steps \(t = 1, \ldots, T\).

Each non-bot user has a state that includes interests, stance, memory, activity, susceptibility, and influence:
\begin{equation}
s_u^t = \{p_u, o_u^t, m_u^t, a_u, \rho_u, \iota_u\}.
\end{equation}
Each content item has a representation:
\begin{equation}
x_c = [z_c, s_c, e_c, q_c, \tau_c, r_c],
\end{equation}
where \(z_c\) is the topic vector, \(s_c\) is stance, \(e_c\) is emotional intensity, \(q_c\) is quality, \(\tau_c\) is the timestamp, and \(r_c\) is risk. Organic users, organic items, and historical interactions are initialized from public recommendation datasets. Target coordinated items are introduced only within the offline evaluation.

Risk is a simulator-known, label-conditioned variable rather than the output of a deployable detector. It is sampled from different target and organic distributions and enters the deterministic agent score, the feedback ranker, and the credibility-aware defenses. This information regime is held fixed for matched comparisons but can affect both behavior APR and later exposure through feedback; consequently, absolute APR values are conditional on it.

At time \(t\), a recommender returns a Top-K list for every non-bot user:
\begin{equation}
R_u^t = \mathrm{TopK}_c \; Score_t(u,c), \quad u \in U_N \cup U_L.
\end{equation}
Non-bot behavior is sampled from action-specific probabilities for click, like, comment, and share:
\begin{equation}
y_{u,c,t}^{a} \sim \mathrm{Bernoulli}(P_a(u,c,t)).
\end{equation}
The behavior probability combines preference match, stance alignment, emotional intensity, content quality, freshness, social heat, influence, susceptibility, and a profile-conditioned agent score:
\begin{align}
P_a(u,c,t) = \sigma(& b_a + \theta_m M_{uc} + \theta_s A_{uc}
 + \theta_e E_c + \theta_q Q_c \nonumber\\
& + \theta_f F_{ct} + \theta_h H_{ct}\rho_u
 + \theta_i \iota_u + \mu G_{uct}).
\end{align}
Here \(M_{uc}\), \(A_{uc}\), \(E_c\), \(Q_c\), \(F_{ct}\), \(H_{ct}\), and \(G_{uct}\) denote match, stance alignment, emotion, quality, freshness, heat, and the profile-conditioned agent score, respectively.
The coordinated policy is represented as:
\begin{equation}
\pi_A = \{r_b, \mu_{sync}, \mu_{sem}, \mu_{mask}, \lambda_{int}, b_{topic}, \kappa_{time}\}.
\end{equation}
Here \(r_b\) is the coordinated-account ratio, \(\mu_{sync}\) controls behavioral synchronization, \(\mu_{sem}\) controls semantic consistency, \(\mu_{mask}\) controls semantic disguise, \(\lambda_{int}\) controls interaction intensity, \(b_{topic}\) controls target-topic bias, and \(\kappa_{time}\) controls temporal dispersion. The evaluation problem is to estimate how \(\pi_A\), the recommender, the data domain, and the defense policy jointly affect target-content penetration among \(U_N \cup U_L\).

\section{CoSimRec Framework}

\subsection{Overview}
CoSimRec couples coordinated interactions, recommender updating, non-bot exposure and response, and ranking intervention in a shared feedback state. Algorithm~\ref{alg:cosimrec} summarizes one episode.

\begin{algorithm}[t]
\caption{CoSimRec Feedback-Loop Simulation}
\label{alg:cosimrec}
\small
\begin{algorithmic}[1]
\REQUIRE User sets \(U_N,U_L,U_B\); organic content \(C_O\); target content \(C_A\); recommender \(R\); coordination policy \(\pi_A\); defense \(D\); horizon \(T\); list size \(K\).
\ENSURE Exposure log, interaction log, APR metrics, heat trajectory, ranking statistics, and CTR.
\STATE Initialize user profiles, content features, target items, and recommender state \(S_0\).
\FOR{\(t=1,\ldots,T\)}
    \STATE Generate coordinated target interactions \(B_t \leftarrow \pi_A(U_B,C_A,t)\).
    \STATE Update feedback state \(S_t \leftarrow Update(S_{t-1},B_t)\).
    \FORALL{\(u \in U_N \cup U_L\)}
        \STATE Produce defended recommendation list \(R_u^t \leftarrow TopK(R,S_t,u,D,K)\).
        \STATE Sample non-bot actions \(Y_u^t\) from the controlled behavior model.
    \ENDFOR
    \STATE Update feedback state \(S_t \leftarrow Update(S_t,\{Y_u^t\}_{u \in U_N \cup U_L})\).
    \STATE Record target exposure, target engagement, heat, ranks, APR, APR\(_B\), and CTR.
\ENDFOR
\STATE \textbf{return} aggregate penetration and utility metrics.
\end{algorithmic}
\end{algorithm}

\subsection{Controlled User-Agent Module}
The recommender selects each non-bot user's list; the behavior model then samples clicks, likes, comments, and shares. This separates target exposure from engagement.

Exposure APR and behavior APR correspond to successive stages of the feedback loop. Exposure APR is computed from served lists before any user action. Behavior APR additionally depends on preference match, susceptibility, content attributes, and response stochasticity. Separate reporting distinguishes broad but weakly engaging exposure from limited exposure followed by concentrated engagement.

The agent design keeps behavioral complexity controllable and inspectable. A rule-based mode uses probabilistic preference features only; profile and memory variants condition responses on user histories; and ablation variants remove heat or risk signals to test whether behavior-level metrics depend on feedback state and content risk. The primary experiments use a pre-specified probabilistic response model conditioned on profiles, memory, candidate content, and dynamic state. These agents provide controlled cross-condition comparisons, not empirically calibrated models of clicking, rating, or listening.

All datasets use the same behavior model. Exposure APR is the stronger primary endpoint because it is read directly from served lists; behavior APR and exposure changes after user feedback are model-conditional sensitivity results. Section~\ref{sec:limitations} discusses the shared response model and target-risk information.

\subsection{Environment Initialization}
CoSimRec initializes users, items, and interactions from public datasets, then adds target items as experimental treatments. This preserves the observed graph or preference structure across matched policies.

Matched trials reuse user profiles, organic candidates, and target items. The coordinated policy and the feedback it induces are the only changing inputs. Retaining target items in the no-attack trial controls for candidate-set expansion, while dataset histories preserve the user, item, and graph structure seen by each recommender.

\begin{table*}[t]
\caption{Datasets and Historical Interaction Signals}
\label{tab:datasets}
\centering
\scriptsize
\begin{tabular}{p{0.18\textwidth}p{0.12\textwidth}p{0.26\textwidth}p{0.34\textwidth}}
\toprule
Dataset & Domain & Historical signal & Evaluation role \\
\midrule
MIND & News & Click and impression histories & News-feed exposure setting \\
MovieLens 1M & Movies & Explicit movie ratings & Main APR and closed-loop LightGCN settings \\
MovieLens latest-small & Movies & Explicit movie ratings & Development-scale LightGCN profile-injection validation \\
LastFM & Music & Artist listening counts & Implicit-feedback preference setting \\
\bottomrule
\end{tabular}
\end{table*}

\subsection{Recommender Feedback Mechanisms}
Random recommendation is the non-feedback control. Popularity ranking responds to accumulated heat, while the feedback ranker combines match, heat, freshness, collaborative feedback, quality, and risk. MF and BPR-MF are latent-factor baselines \cite{koren2009matrix,rendle2009bpr}. LightGCN uses trainable embeddings, normalized bipartite propagation, and BPR optimization \cite{he2020lightgcn}. We first evaluate it with standard ASR and held-out ranking utility, then place the trained model in the APR loop to measure non-bot exposure and response after graph updates.

The recommender suite covers distinct propagation paths. Random ranking estimates chance exposure; popularity ranking measures direct accumulation; and the feedback ranker combines global heat with user-specific signals. MF and BPR-MF introduce latent preference objectives, whereas LightGCN tests transfer through user--item connectivity. Comparing them tests which ranking signals carry coordinated interactions into non-bot lists.

The CoSimRec feedback ranker uses:
\begin{align}
Score_t(u,c) ={}& \alpha M_{uc} + \beta H_{ct} + \gamma F_{ct}
 + \delta CF_{uct} \nonumber\\
& + QBonus(c) - \lambda Risk(c).
\end{align}
Here \(CF_{uct}\) denotes collaborative feedback and \(QBonus(c)\) denotes a quality bonus. This transparent scoring function makes the contributions of heat, user-content match, freshness, collaborative feedback, quality, and risk directly inspectable.

\subsection{Coordination Policies}
No attack is the baseline. Random bots distribute target interactions across accounts and time, whereas master-worker coordination concentrates accounts on shared targets within narrower windows. These simulator-policy labels are analytical and do not imply that corresponding real accounts are automated or deceptive. Their 30-step budgets are 552 and 1,120 target events, respectively; because both volume and temporal concentration vary, comparisons identify the total effect of each policy package rather than an isolated synchronization effect.

\subsection{APR Metric Family}
Exposure APR measures the fraction of non-bot recommendation capacity occupied by target coordinated content:
\begin{equation}
\mathrm{APR}_E@K =
\frac{\sum_t \sum_{u \in U_N \cup U_L} \sum_{c \in C_A} X_{u,c,t}^K}
{\sum_t \sum_{u \in U_N \cup U_L} \sum_{c \in C} X_{u,c,t}^K + \epsilon}.
\end{equation}
Here \(X_{u,c,t}^K=1\) if item \(c\) appears in the Top-K list of user \(u\) at time \(t\). Behavior APR measures the fraction of non-bot engagement directed to target coordinated content:
\begin{equation}
\mathrm{APR}_B =
\frac{\sum_t \sum_{u \in U_N \cup U_L} \sum_{c \in C_A} Engage(u,c,t)}
{\sum_t \sum_{u \in U_N \cup U_L} \sum_{c \in C} Engage(u,c,t) + \epsilon}.
\end{equation}
The engagement weight is:
\begin{equation}
Engage = Click + Like + 2 Comment + 3 Share.
\end{equation}
APR-Lift compares each coordinated policy with a matched no-attack baseline:
\begin{equation}
\mathrm{APR\text{-}Lift} = \mathrm{APR}_E(\pi_A) - \mathrm{APR}_E(NoAttack).
\end{equation}
Penetration gain normalizes additional non-bot target exposure by the coordinated interaction budget:
\begin{equation}
PenetrationGain =
\frac{E_N(C_A | \pi_A) - E_N(C_A | NoAttack)}
{Budget(\pi_A) + \epsilon}.
\end{equation}
APR is computed only for ordinary users and opinion leaders. Coordinated-account interactions are treated as budget and feedback input, not as successful penetration.

APR complements rather than replaces attack-side metrics. ASR and target-rank improvement measure model-level promotion, final target hit rate measures recipient coverage at one checkpoint, and \(\mathrm{APR}_E\) measures the target share of all non-bot slots over the loop. Reporting these stages together can reveal promotion that does not transfer to recipients, as in the disconnected LightGCN conditions. Penetration gain additionally assumes that coordinated events are cost-homogeneous; heterogeneous account, timing, and action costs would require a weighted budget.

\subsection{Ranking Interventions}
CoSimRec evaluates ranking interventions as score-level modifications:
\begin{align}
Score'_t(u,c) ={}& Score_t(u,c) - \lambda_1 Heat(c)
 - \lambda_2 SynRisk_t(c) \nonumber\\
& - \lambda_3 SemDup(c) - \lambda_4 CredRisk(c) \nonumber\\
& + Diversity(c,R_u^t).
\end{align}
The evaluated conditions are no defense, heat down-weighting, diversity reranking, semantic penalty, synchronization penalty, credibility penalty, and their combination. The main defense outcome is APR reduction:
\begin{equation}
Reduction(APR) =
\frac{APR_{NoDefense} - APR_{Defense}}
{APR_{NoDefense} + \epsilon}.
\end{equation}
The utility proxy is CTR reduction relative to the no-defense condition. A negative CTR reduction means that CTR increased slightly under the defense.

\section{Experimental Protocol}
Primary APR experiments use ten paired seeds, 200 agents, 1000 items including 40 targets, 30 steps, and Top-10 lists. We compare no attack, random bots, and master-worker coordination under random, popularity, and feedback ranking. The risk-blind primary protocol withholds simulator-known risk from ranking and response; a factorial ablation varies this access. APR-Lift uses matched baselines, with paired bootstrap intervals and exact sign-flip tests. Appendix~\ref{sec:supp-config} and the reproducibility artifact provide the remaining settings.

\section{Results}

\begin{figure*}[htbp]
\centering
\includegraphics[width=0.78\textwidth]{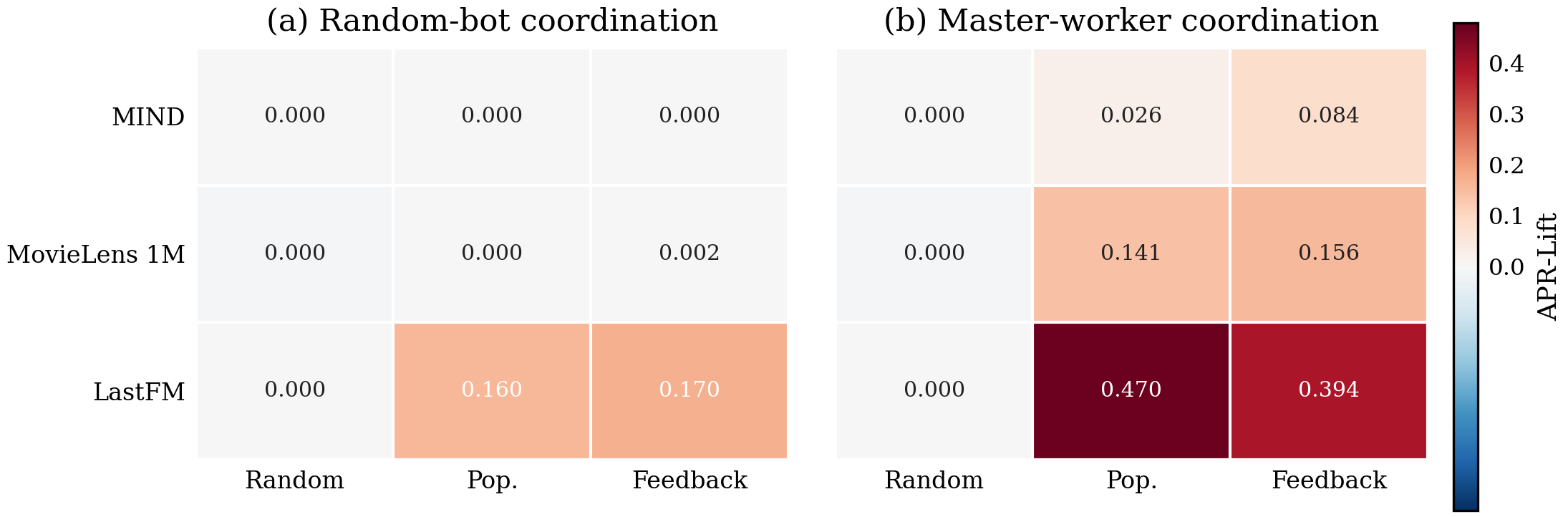}\\[-0.2em]
\includegraphics[width=0.78\textwidth]{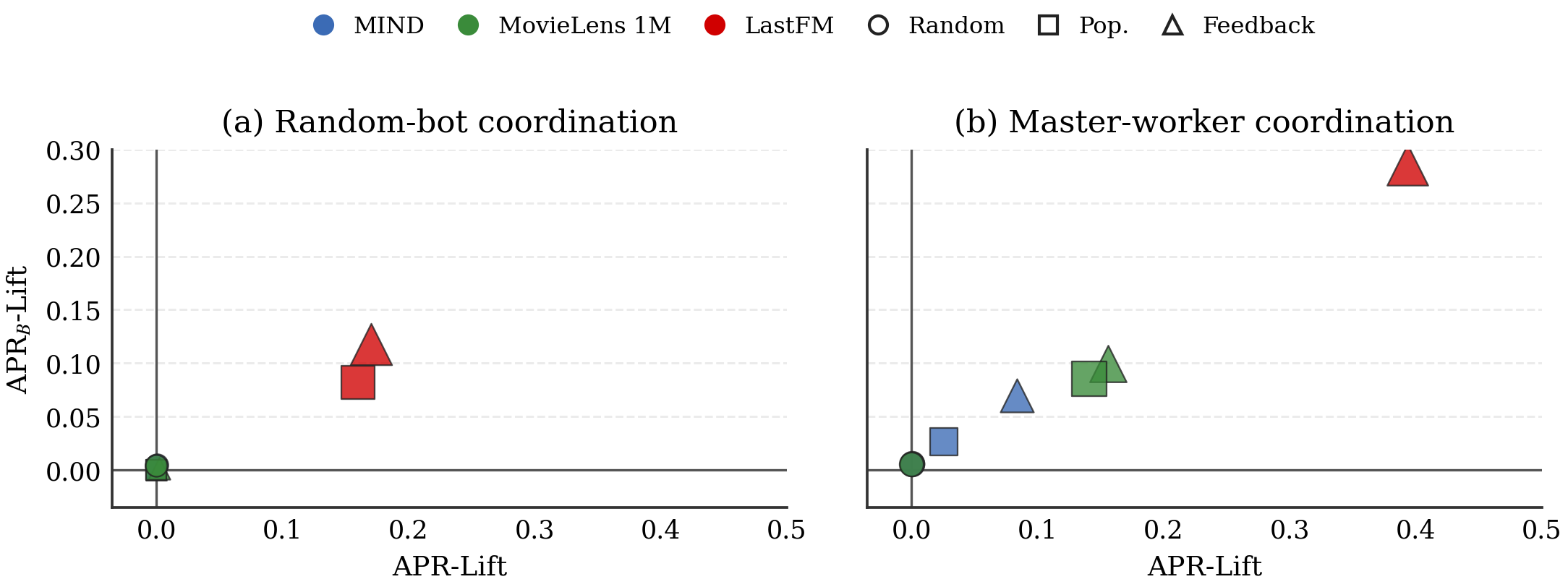}
\caption{Risk-blind main penetration evidence from the ten-seed primary artifact. Top: APR-Lift heat map across datasets, recommenders, and coordination policies. Bottom: exposure--behavior coupling after matched-baseline subtraction, where the \(x\)-axis denotes APR-Lift, the \(y\)-axis denotes APR\(_B\)-Lift, color denotes dataset, marker shape denotes recommender, and marker size denotes peak heat.}
\label{fig:main-evidence}
\end{figure*}

\begin{table*}[t]
\caption{Master-Worker APR-Lift Robustness Across Ten Seeds}
\label{tab:main-stats}
\centering
\scriptsize
\begin{tabular}{llccc}
\toprule
Dataset & Recommender & APR-Lift mean [bootstrap 95\% CI] & Exact BH \(p\) & Direction \\
\midrule
MIND & Random & 0.0001 [-0.0005, 0.0007] & 0.9649 & None \\
MIND & Popularity & 0.0258 [0.0218, 0.0297] & 0.0032 & Positive \\
MIND & Feedback & 0.0838 [0.0719, 0.0968] & 0.0032 & Positive \\
MovieLens & Random & -0.0003 [-0.0009, 0.0003] & 0.4506 & None \\
MovieLens & Popularity & 0.1407 [0.1294, 0.1508] & 0.0032 & Positive \\
MovieLens & Feedback & 0.1556 [0.1324, 0.1754] & 0.0032 & Positive \\
LastFM & Random & 0.0002 [-0.0007, 0.0011] & 0.8300 & None \\
LastFM & Popularity & 0.4702 [0.4562, 0.4855] & 0.0032 & Positive \\
LastFM & Feedback & 0.3936 [0.3738, 0.4147] & 0.0032 & Positive \\
\bottomrule
\end{tabular}
\end{table*}

\begin{table*}[t]
\caption{BPR-LightGCN profile-injection results on MovieLens latest-small across five seeds (mean \(\pm\) seed-level SD). Coordinated profiles connect one long-tail target to 20 popular filler items.}
\label{tab:lightgcn-validation}
\centering
\scriptsize
\begin{tabular}{lccccc}
\toprule
Injection & Target ASR@20 & Rank improvement & Recall@20 & NDCG@20 & $\Delta$NDCG@20 \\
\midrule
Clean (0\%) & \(0.0000 \pm 0.0000\) & \(0.0 \pm 0.0\) & \(0.0701 \pm 0.0022\) & \(0.0293 \pm 0.0007\) & \(0.0000 \pm 0.0000\) \\
Bandwagon (5\%) & \(0.0000 \pm 0.0000\) & \(1252.7 \pm 54.3\) & \(0.0741 \pm 0.0022\) & \(0.0304 \pm 0.0002\) & \(0.0010 \pm 0.0007\) \\
Bandwagon (10\%) & \(0.0549 \pm 0.0292\) & \(1374.6 \pm 67.2\) & \(0.0708 \pm 0.0015\) & \(0.0292 \pm 0.0004\) & \(-0.0001 \pm 0.0006\) \\
Bandwagon (20\%) & \(0.9700 \pm 0.0670\) & \(1418.0 \pm 72.5\) & \(0.0701 \pm 0.0034\) & \(0.0283 \pm 0.0010\) & \(-0.0010 \pm 0.0016\) \\
\bottomrule
\end{tabular}
\end{table*}

\subsection{Multi-Target LightGCN Stress Test}
We extend the closed-loop test to nine explicit MovieLens 1M targets: three long-tail, three medium-popularity, and three popular items. Each target is evaluated over five seeds, six injection ratios (5\%--30\%), and matched no-attack, target-only, and connected-bandwagon conditions. Unlike the synthetic long-tail validation, an explicit target can already have historical graph edges; target-only serves as a \emph{no-filler injection} control, not as evidence of disconnection.

\begin{figure*}[t]
\centering
\includegraphics[width=0.92\textwidth]{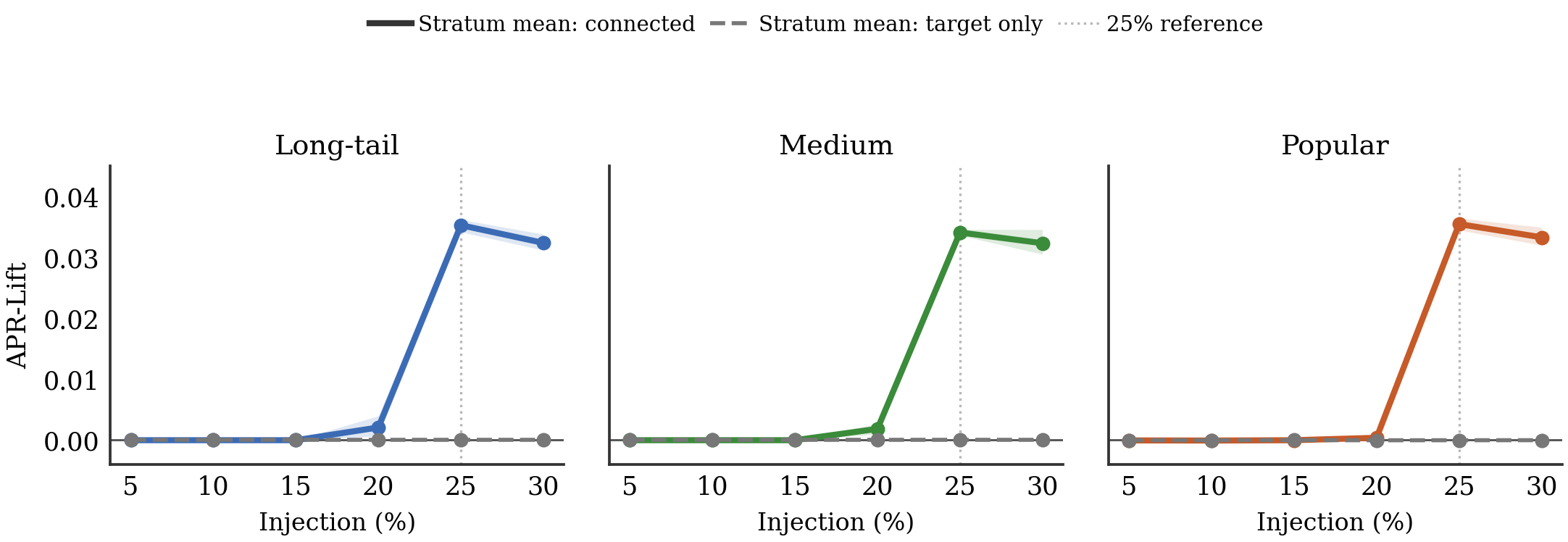}
\caption{MovieLens 1M multi-target LightGCN stress test. Each panel aggregates three explicit targets from one degree stratum. Solid colored lines and shaded bands show the stratum mean and seed-bootstrap 95\% interval for connected-bandwagon profiles; gray dashed lines show the stratum mean for the target-only (no-filler) control. The vertical dotted line marks 25\% injection. Target-level results are provided in the accompanying artifact.}
\label{fig:multitarget-lightgcn}
\end{figure*}

Fig.~\ref{fig:multitarget-lightgcn} shows every target and its seed-bootstrap interval. Connected bandwagon profiles produce essentially zero APR-Lift through 15\% injection for every target stratum. At 20\%, the stratum means are small and heterogeneous (0.0021, 0.0019, and 0.0004 for long-tail, medium, and popular targets). At 25\%, the corresponding means increase to 0.0353, 0.0341, and 0.0355; all 45 target--seed estimates are positive. At 30\%, the stratum means remain positive (0.0324, 0.0324, and 0.0333). Target-only profiles remain near zero at all ratios. This is a five-seed screening result under one filler strategy and retraining schedule, rather than an estimate of a universal injection threshold.

\subsection{Penetration Across Ranking Mechanisms}
Fig.~\ref{fig:main-evidence} reports APR-Lift by dataset, recommender, and coordination policy. Under master-worker coordination, random recommendation stays close to 0.000 APR-Lift on MIND, MovieLens, and LastFM. The intervals for MIND and MovieLens include zero; LastFM has a statistically detectable but negligible negative lift. Because target items are present in these random-control runs, their near-zero lift separates target availability from the positive effects observed under feedback-sensitive ranking. The magnitude of those effects varies by dataset and recommender.

The master-worker policy package produces more exposure than the lower-budget random-bot package under feedback-sensitive ranking. LastFM reaches APR-Lift of 0.470 under popularity and 0.394 under feedback; MovieLens reaches 0.141 and 0.156. Because the policies differ in realized event volume as well as temporal concentration, this contrast does not identify a synchronization effect. Table~\ref{tab:lightgcn-validation} verifies target-rank promotion in the development-scale benchmark, while Fig.~\ref{fig:multitarget-lightgcn} provides the recipient-side closed-loop evidence across nine explicit MovieLens targets.

Near-zero random-recommender controls and larger lifts under feedback-sensitive ranking suggest that target availability alone does not account for the results. Within the controlled simulator, temporal or graph structure may become a reusable ranking signal; this interpretation is not a causal estimate for deployed communities.

The lower panel of Fig.~\ref{fig:main-evidence} relates exposure APR-Lift to behavior APR lift under the same matched baseline. The two measures are related but not interchangeable. Recommendation controls whether target items are shown, while user-agent scoring controls whether exposed non-bot users engage with them. Marker size reports peak heat, helping distinguish high-exposure cases from high-engagement cases.

Because exposure APR is computed directly from served lists, it does not depend on the sampled action for an exposed item. The response model affects behavior APR and later feedback, but cannot generate engagement for an item absent from the list. Similar exposure with different behavior APR thus indicates variation in conditional response rather than in initial ranking reach.

\subsection{Risk-Information and Response Sensitivity}
The primary results are risk-blind: neither the feedback ranker nor the response model receives simulator-known target risk. A factorial 2$\times$2 ablation independently restores risk to each component. In the fully risk-blind feedback condition, APR-Lift is 0.0838 on MIND, 0.1556 on MovieLens, and 0.3936 on LastFM. Giving risk to the feedback ranker reduces APR by 0.0445, 0.0591, and 0.0642, respectively (paired bootstrap intervals all exclude zero). Thus, the main effect is not produced by a ranker that is told the target label; within this simulator, risk information suppresses rather than creates exposure.

For MIND, we additionally fit a click surrogate using a temporal split (November 9--12, 2019 training; November 13 validation; November 14 held-out test). After validation-set Platt calibration, the held-out AUROC is 0.5925, AUPRC is 0.0581, Brier score is 0.0367, and 15-bin ECE is 0.0008. Substituting this calibrated click-only response in the closed loop raises feedback master-worker exposure APR-Lift from 0.0853 to 0.1020. This is a sensitivity analysis with moderate discrimination, not validation that the simulator recovers human behavior or that the same response model applies to MovieLens or LastFM.

Socially, this distinction separates visibility from reception. Exposure APR measures an opportunity to encounter target content, whereas behavior APR measures conditional uptake under the specified agent model. A high exposure APR marks a shift in attention allocation, not belief change, persuasion, endorsement, or offline action.

\subsection{Seed-Level Statistical Tests}
For all six popularity and feedback conditions, the bootstrap interval excludes zero and the seed-level exhaustive two-sided sign-flip test remains significant after FDR correction. With ten seeds, the minimum attainable unadjusted exact p value is \(2/2^{10}=0.001953\); the six positive APR-Lift tests have BH-adjusted \(p=0.0032\). All three random-control intervals include zero and their adjusted exact p values exceed 0.45. None of the random controls has positive penetration, whereas the largest positive effects occur under LastFM popularity and feedback ranking. The LightGCN results add a graph-specific boundary: low-ratio connected profiles do not penetrate, 20\% is heterogeneous, and 25\%--30\% produces stable positive APR in the nine-target five-seed screening.

\subsection{Population-Scale Sensitivity}
\begin{figure*}[htbp]
\centering
\includegraphics[width=0.90\textwidth]{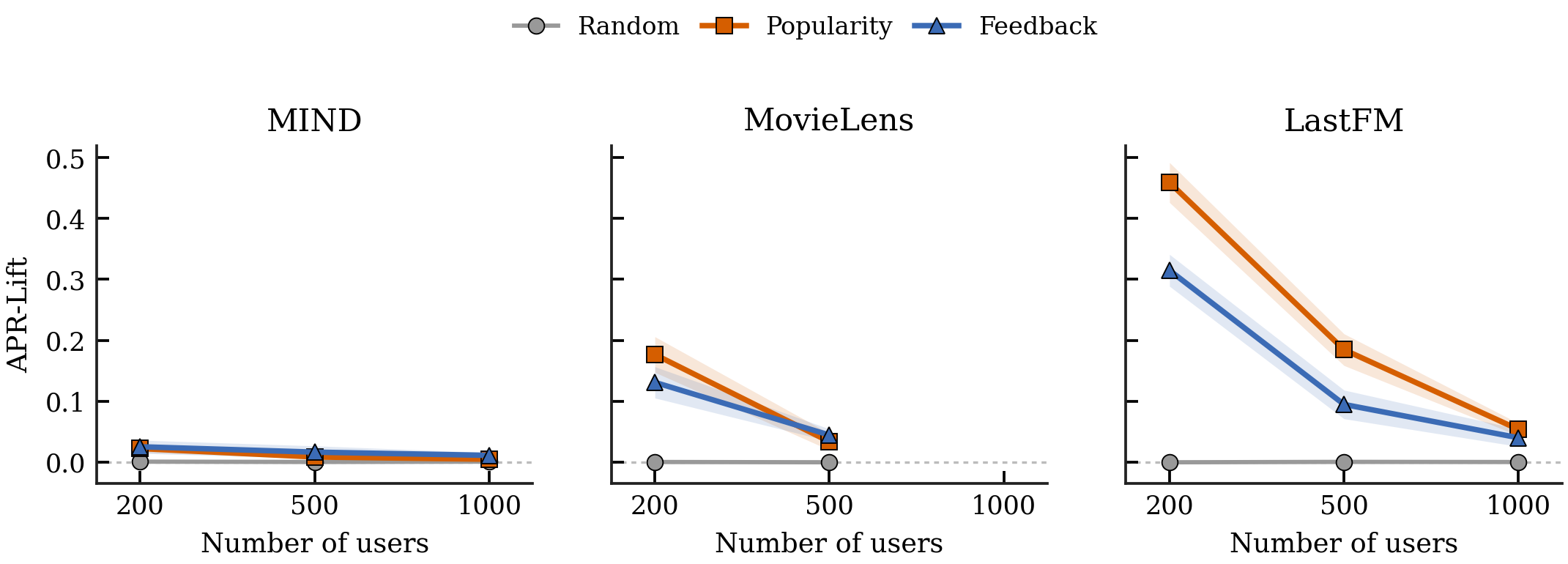}
\caption{Scale-up APR-Lift curves under master-worker coordination. Each panel corresponds to one dataset, lines denote recommenders, and shaded bands denote the seed-level standard deviation across five seeds.}
\label{fig:scale-up}
\end{figure*}

Fig.~\ref{fig:scale-up} reports master-worker APR-Lift for 200, 500, and 1000 users with 1000 items, 30 steps, Top-10 ranking, and five seeds. MovieLens includes up to 500 eligible users.

Random recommendation remains near zero at every scale. On LastFM, popularity and feedback APR-Lift decline as a fixed attack budget is spread over more non-bot exposure.

The decline under scale-up follows the fixed-budget design. Coordinated accounts and target interactions remain constant as the number of non-bot recommendation slots increases, reducing target exposure as a share of the denominator. The experiment characterizes fixed-budget sensitivity; it does not imply that vulnerability decreases when attacker resources scale with the user population.

Penetration is also relational: the same coordination budget can occupy a large or small share of audience attention depending on the size of the exposure environment. Resource concentration and audience scale should be reported together when comparing social-system vulnerability.

\subsection{Ranking-Intervention Trade-Offs}
\begin{figure*}[htbp]
\centering
\includegraphics[width=0.84\textwidth]{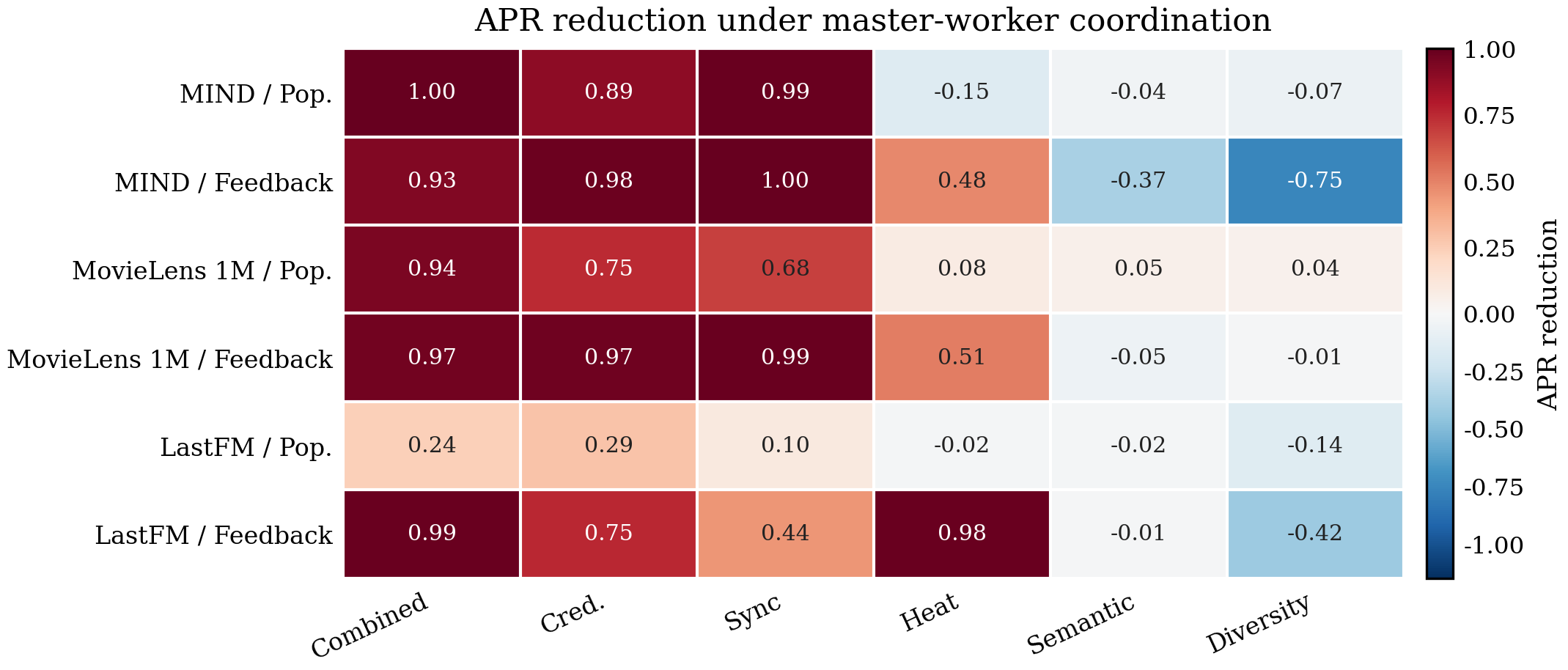}
\caption{Ten-seed APR-reduction heat map for ranking interventions under master-worker coordination. Rows denote the six dataset--recommender settings, columns denote defenses, and cell values denote mean APR reduction relative to no defense. Positive values reduce penetration; negative values increase it. The risk-blind ranking protocol is used; interval estimates are reported in the supplementary artifact.}
\label{fig:defense-heatmap}
\end{figure*}

Fig.~\ref{fig:defense-heatmap} reports APR reduction. Synchronization penalties reduce APR in every case; semantic penalties and diversity have negative means. Oracle credibility and combined policies exceed 0.98 in several feedback-ranking cases. The ten-seed estimates preserve this ordering.

Defense performance follows the signal modified by each intervention. Synchronization penalties act on the temporal concentration induced by master-worker coordination. Heat down-weighting also affects organically popular items and consequently shows greater variance. Diversity and semantic reranking can change the competing candidate set without reducing target scores, which occasionally increases the target share. Credibility penalties use simulator labels and represent an oracle condition rather than an implementable risk estimator.

These interventions also act at different governance loci. Synchronization penalties address the process by which visibility is produced, whereas semantic penalties and diversity reranking primarily alter content composition. Their unequal APR effects show that a policy can change what a list looks like without changing whether coordinated activity captures attention.

Descriptively, synchronization-aware ranking has the largest mean APR reduction among the non-oracle interventions. Its interval is positive in all six dataset--recommender settings (the smallest lower bound is 0.079). Credibility-aware and combined policies produce larger reductions but rely on attack-label-conditioned risk. Heat down-weighting has high variance across settings, while semantic penalty and diversity often yield zero or negative APR reduction. These ten-seed results improve precision but do not establish deployment superiority because the synchrony and credibility signals remain simulated or oracle.

\subsection{Latent-Factor Baselines}
MF and BPR-MF have low APR-Lift in the tested MIND and MovieLens conditions. On LastFM, MF reaches master-worker APR-Lift of 0.2588, while BPR-MF reaches 0.0148. Sensitivity varies by objective, domain, and coordination policy.

MF and BPR-MF should not be treated as a single robustness class. Their objectives weight observed and unobserved interactions differently, producing different target updates from the same coordinated history. The LastFM results further show that low APR on one domain does not establish model-level resistance independent of the interaction distribution.

\subsection{Behavioral-Model Ablation}
Agent ablations separate the contribution of individual behavioral components. Heat and risk components affect behavior-level metrics and agent-score diagnostics. Removing risk raises target-content scores and behavior APR, whereas removing heat changes target scores and engagement patterns. Exposure APR is dominated by recommender ranking and coordinated feedback state. These ablations measure internal sensitivity; they do not test correspondence with observed human behavior.

The LLM sensitivity tests replace the deterministic candidate-scoring component inside the probabilistic response model while retaining the ranking and action-sampling protocol. Moderate rank agreement shows that the two scorers order candidate actions differently, whereas feedback APR-Lift remains near zero in the tested MIND condition. This comparison quantifies sensitivity to the two score specifications but does not validate either as a model of human behavior.

\section{Discussion}

\subsection{Mechanism and Sociotechnical Interpretation}
Coordinated activity is not universally effective: APR-Lift varies by dataset and recommender, so the results identify a conditional mechanism rather than a general vulnerability. The LightGCN studies trace that mechanism from disconnected interactions, through connected target-rank promotion, to non-bot exposure during retraining. The observed 20\%--30\% transition indicates that graph reachability is necessary but not sufficient in this setup; it is not a general threshold.

The results support an affordance-matching interpretation: coordination gains influence when its temporal or graph structure matches signals rewarded by ranking. The algorithm thereby acts as a gatekeeper between organized activity and audience visibility \cite{gillespie2014relevance,napoli2014automated}. This resembles cumulative advantage in cultural markets \cite{salganik2006experimental}, but the analogy is interpretive because CoSimRec changes recommendation feedback rather than users' perception of social proof.

APR separates manufactured visibility from modeled uptake and from unmeasured persuasion or collective effects. Exposure can change opportunities for attention without changing attitudes; future work can examine unequal reach through group-stratified APR.

\subsection{Dataset Differences}
Historical interactions shape penetration. Repeated artist preferences may explain LastFM's sensitivity to popularity and feedback ranking. MovieLens graph attacks require shared filler items, while MIND's topical histories make targets harder to align with users. These explanations remain hypotheses rather than tested causal mechanisms.

Domain-specific interaction structure changes the available feedback signal. Repeated music consumption supports persistent artist-level heat, movie ratings yield a relatively stable preference graph, and news relevance decays rapidly. These properties alter freshness, collaborative overlap, and the rate of heat accumulation. Since the response specification is shared across datasets, the observed differences characterize dataset--recommender interactions rather than platform-level prevalence.

Sociologically, the three datasets are not interchangeable containers. Listening, rating, and clicking encode different temporal rhythms and meanings of participation, and each recommender institutionalizes those traces differently. Cross-domain variation informs the interaction between social practice and technical design, but it is not direct evidence about the prevalence or impact of manipulation in music, movie, or news publics.

\subsection{Defense Implications}
Synchronization penalties produce larger descriptive APR reductions than generic diversification and use temporal or account-level signals that may be observable without target labels. However, the experiments assume error-free synchronization signals, while credibility and combined policies use oracle labels. Deployment studies must therefore evaluate calibration, ranking utility, false positives, and detection delay alongside APR reduction.

Coordination signals should support graded friction, audit, or corroboration rather than automatic suppression because synchronized communication also enables civic mobilization, fan participation, and emergency response \cite{bennett2012connective}. Coordination detectors can produce false positives on out-of-distribution legitimate activity \cite{vargas2020detection}; evaluations should examine appeals and error distributions across communities.

\section{Limitations, Ethics, and Threats to Validity}
\label{sec:limitations}
\textbf{Scale and external validity.} CoSimRec provides controlled offline comparisons rather than estimates of attack prevalence on deployed platforms. The main experiments use 200 users, 1000 items, 30 steps, Top-10 ranking, ten seeds, and fixed coordinated budgets. Larger catalogs, longer horizons, adaptive campaigns, or resources that scale with the population may produce different effects.

\textbf{Target and LightGCN design.} Targets are controlled treatments within historical logs and abstract the semantics, timing, and production choices of real campaigns. The LightGCN extension covers nine MovieLens targets but only five seeds, one filler policy, one catalog sample, and a bounded retraining schedule; its 25\% transition region does not generalize beyond these conditions. Its affinity-based preference NDCG is distinct from held-out benchmark accuracy.

\textbf{Behavior and information assumptions.} Behavior APR reflects a specified response model rather than observed human uptake. The short-horizon LLM study is a sensitivity check, and the MIND surrogate has moderate discrimination with no parallel calibration for MovieLens or LastFM. The primary protocol is risk-blind, but credibility-based defenses remain oracle upper bounds requiring independently estimated and calibrated risk signals.

\textbf{Social and ethical boundary.} APR measures distributional reach, not persuasion, endorsement, attitude change, or social harm. User roles and the terms ``target'' and ``attack'' are analytical constructs; legitimate coordination may be misclassified, and historical logs are not a neutral population baseline. The interventions act only at ranking time. All experiments use public, de-identified secondary data offline, without recruited participants, new human-subject data, platform APIs, or live users.

\section{Conclusion}
CoSimRec connects attacker-side promotion to recipient-side visibility through a matched offline feedback-loop protocol and the APR metric family. Under risk-blind conditions, random ranking shows no positive master-worker APR-Lift, whereas popularity and feedback ranking do, reaching 0.4702 on LastFM. In the MovieLens 1M stress test, connected profiles produce APR-Lift of 0.034--0.036 at 25\% injection across all three target-popularity strata, while no-filler profiles remain near zero.

These results show a computational possibility under bounded response, attack, and retraining assumptions, not the prevalence, persuasion, harm, or defense effectiveness of deployed manipulation. Within this boundary, penetration depends on the fit among coordinated behavior, ranking signals, graph and temporal structure, and audience scale. APR isolates that mechanism without treating target-rank promotion as evidence of audience exposure.

\section*{Data and Code Availability}
The code, configurations, and instructions required to reproduce the experiments and reported results are publicly available at \url{https://github.com/luan-luan-nan/CoSimRec}. Public datasets remain with their original providers and are not redistributed.

\bibliographystyle{IEEEtran}
\bibliography{references}

\clearpage
\appendices
\section{Supplementary Experimental Details}
\label{sec:supp-config}

\subsection{Multi-Target BPR-LightGCN Configuration}
The multi-target stress test uses nine explicit MovieLens 1M targets (three each from long-tail, medium, and popular degree strata), five seeds (42--46), and fake-profile ratios of 5\%, 10\%, 15\%, 20\%, 25\%, and 30\%. Each target--seed pair is run with no attack, target-only, and connected-bandwagon profiles, producing 810 paired records. Connected profiles use 20 popular fillers. Because explicit targets can have observed historical edges, target-only is interpreted only as a no-filler injection control. Final target rank is measured before the seen-item mask used for recommendation; full training and update parameters are provided in the reproducibility artifact.

\subsection{Risk-Information and MIND Calibration Configuration}
The primary APR protocol sets both \texttt{ranker\_uses\_risk} and \texttt{response\_uses\_risk} to false. The factorial ablation crosses these switches, with ten paired seeds (42--51), feedback ranking, and no-attack versus master-worker conditions. This isolates whether simulator-known target risk enters the response loop, the ranker, both, or neither.

For MIND calibration, behavior logs from November 9--12, 2019 train a logistic click surrogate using only training-derived user profiles and counts. November 13 is used for Platt calibration and November 14 for held-out evaluation (30,270 behavior rows and 1,222,429 candidate impressions). The calibrated alternative response mode produces clicks only; it is consequently compared on exposure APR, late-round APR, and reach, not behavior APR.

\subsection{Controlled Agent Configuration}
Two five-seed MIND sensitivity analyses replace the deterministic candidate score inside the probabilistic response model with scores from \texttt{gpt-5-nano-2025-08-07}. Both use seeds 42--46, temperature 0, three feedback steps, strict parsing with no fallback, and prompts containing the profile, recent memory, candidate attributes, heat, and simulator-known risk. A deterministic 10\% model-agent sample yields 430 model-backed scores and mean Spearman agreement 0.478 with the fixed scorer. Routing all opinion leaders, and only opinion leaders, through the model yields 600 model-backed scores and mean agreement 0.443 across 16 nonconstant comparisons; feedback APR-Lift is 0.0019 with a 95\% interval of [-0.0021, 0.0058]. Prompt hashes, cache provenance, sampling rules, and per-seed manifests are included in the artifact.

\subsection{Defense Configuration}
Defense experiments use ten seeds (42--51) with popularity and feedback ranking. We test heat down-weighting, diversity, semantic, synchronization, credibility, and combined penalties. MovieLens and LastFM use master-worker coordination; MIND also includes random bots. Outcomes are APR, APR\(_B\), and CTR reduction.

\subsection{Statistical Tests}
Statistical tests use seed-level measurements from the experiments. For main attacks, APR-Lift and APR\(_B\)-Lift are tested against zero across ten seeds. CTR and preference NDCG are tested as paired per-seed deltas against the corresponding no-attack baseline. For defenses, APR reduction, APR\(_B\) reduction, and CTR reduction are summarized across ten seeds under master-worker coordination. Main-attack uncertainty uses 20,000 paired-seed bootstrap resamples; exact two-sided sign-flip p values enumerate all \(2^{10}=1024\) sign assignments and are then Benjamini--Hochberg adjusted within the stated family. The multi-target LightGCN results are explicitly descriptive five-seed screening estimates.

Attack and defense runs are paired by seed with their reference conditions. Bootstrap intervals quantify uncertainty in the mean paired effect, and the sign-flip test provides nonparametric inference for the ten-seed attack comparisons. The primary-sweep BH family contains 54 tests: three datasets, three recommenders, two attack policies, and three metrics (APR-Lift, APR\(_B\)-Lift, and CTR delta). The defense family contains 108 tests: six dataset--recommender settings, six interventions, and three metrics. Defense conclusions emphasize paired reductions, intervals, and the observability of their input signals rather than significance counts alone.

\end{document}